\documentclass[12pt]{article}
\usepackage{graphicx}
\usepackage[cp1251]{inputenc}
\usepackage{epsfig}
\usepackage[english]{babel}

\date{}

\begin{document}
\setcounter{page}{1}
\pagestyle{plain}

\title{\bf{Local moment formation in bilayer graphene}}

\author{Yawar Mohammadi$^1$\thanks{Corresponding author. Tel./fax: +98 831 427
4569, Tel: +98 831 427 4569. E-mail address:
yawar.mohammadi@gmail.com} , Rostam Moradian$^{2,3}$}
\maketitle{\centerline{$^1$Department of Physics, Islamic Azad
University, Kermanshah Branch, Kermanshah, Iran}
\maketitle{\centerline{$^2$Department of Physics, Razi University,
Kermanshah, Iran} \maketitle{\centerline{$^3$Nano Science and Nano
Technology Research Center, Razi University, Kermanshah, Iran}

\begin{abstract}

The local properties of bilayer graphene (BLG) due to the spatial
inhomogeneity of its sublattices are of interest. We apply
Anderson impurity model to consider the local moment formation on
a magnetic impurity which could be adsorbed on different
sublattices of BLG. We find different features for the impurity
magnetization when it is adsorbed A and B sublattices. The
impurity adsorbed on A sublattice can magnetize even when the
impurity level is above the Fermi level and the on-site coulomb
energy is very small. But when the impurity is adsorbed on B
sublattice the magnetization is possible for limited values of the
impurity level and the on-site coulomb energy. This is due to
different local density of the low energy states at A and B
sublattices which originates from their spatial inhomogeneity.
Also we show that electrical controlling the magnetization of
adatoms besides it's inhomogeneity in BLG allow for possibility of
using BLG in spintronic devices with higher potential than
graphene.
\end{abstract}


\vspace{0.5cm}

Green's function; D. Local moment.
%
\section{Introduction}
\label{sec:1}

Single layer graphene (SLG), a single layer of carbon atoms
arranged in a honeycomb lattice, has attracted many experimental
and theoretical efforts in the last decade. These efforts result
in discovery of many unusual properties\cite{Castro Neto1,Peres1}
which are due to the massless chiral Dirac nature of its charge
carriers. One of the attractive topics in SLG is study of the
local moment formation on a magnetic impurity adsorbed on a
SLG\cite{Uchoa,Cornaglia,Mao,Li,Wehling}. This is motivated by the
use of the scanning tunnelling microscope to control position of
an impurity adsorbed on a two-dimensional open
surface\cite{Eigler,Brar}. Uchoa $et$ $al.$ used Anderson
model\cite{Anderson} to investigate the necessary conditions for
the formation of the local moment on a magnetic impurity adsorbed
on a SLG. They found that the impurity adsorbed on SLG magnetizes
even when the energy level of the impurity is above the Fermi
energy and the on-site coulomb interaction is small. Also they
showed that one can control the magnetic
 moment formation via an external electric field. These behaviors are
different from the impurity magnetization in normal metals.
This method could be expanded to investigate the impurity magnetization in BLG.

Bilayer graphene (BLG), composed of two layers of graphene with
strong interlayer tunnelling, similar to the ordinary
two-dimensional electron gas (2DEG) has parabolic band structure.
Despite of this similarity, BLG shows unusual properties which are
not observed in 2DEG\cite{Das Sarma1,McCann1}. Due to this feature
of BLG, the study of impurity magnetization in
BLG can be also attractive. Recently several groups have
investigated the possibility of the local moment formation on a magnetic impurity
adsorbed on BLG\cite{Ding,Killi,Sun}. To scrutinize this topic
more, we study it with special emphasis on different features of
the local moment formation when the impurity is adsorbed on
different sublattices of BLG. We also address the potentiality of
BLG doped with the magnetic impurities for spintronics. The paper
is organized as follows. The model Hamiltonian and the details of our
calculations are presented in the section II. In the section III
we discuss our numerical results. Finally we end this paper by
summary and conclusions in the section IV.

\section{Model Hamiltonian}
\label{sec:2}

We apply Anderson model to study the necessary conditions for the
local moment formation on an impurity adsorbed on the top of
different sublattices of BLG. The total Hamiltonian of a BLG which
has adsorbed a magnetic impurity is
\begin{eqnarray}
H_{T}=H_{BLG}+H_{imp}+H_{V}
,\label{eq:01}
\end{eqnarray}
where the Hamiltonian of pure BLG in the
nearest neighbor tight-binding approximation is
\begin{eqnarray}
H_{BLG}=-t\sum_{m=1}^{2}\sum_{<ij>,\sigma}[a_{m\sigma}^{\dag}(\mathbf{R}_{i})b_{m\sigma}(\mathbf{R}_{j})+
b_{m\sigma}^{\dag}(\mathbf{R}_{j})a_{m\sigma}(\mathbf{R}_{i})]+
\gamma\sum_{i,\sigma}[a_{1\sigma}^{\dag}(\mathbf{R}_{i})b_{2\sigma}(\mathbf{R}_{i})+
b_{2\sigma}^{\dag}(\mathbf{R}_{i})a_{1\sigma}(\mathbf{R}_{i})]
,\label{eq:02}
\end{eqnarray}
where
$a_{m\sigma}^{\dag}(\mathbf{R}_{i})(a_{m\sigma}(\mathbf{R}_{i}))$
creates(annihilates) an electron with spin $\sigma$ at $A$
sublattice in site $i$ of m-th layer. $t=2.7$ eV and $\gamma=0.4$
eV present the nearest neighbor intralayer
($A_{1}\longleftrightarrow B_{1}$ or $A_{2}\longleftrightarrow
B_{2}$) and interlayer ($A_{1}\longleftrightarrow B_{2}$) hopping
energies respectively. Fig. \ref{fig:01} shows a BLG lattice. One
can diagonalize the momentum dependent Hamiltonian of pure BLG,
\begin{eqnarray}
H_{BLG}=-t\sum_{m=1}^{2}\sum_{\mathbf{k}\sigma}[\phi(\mathbf{k})a_{m\mathbf{k}\sigma}^{\dag}
b_{m\mathbf{k}\sigma}+
\phi^{\ast}(\mathbf{k})b_{m\mathbf{k}\sigma}^{\dag}a_{m\mathbf{k}\sigma}]+
\gamma\sum_{\mathbf{k}\sigma}[a_{1\mathbf{k}\sigma}^{\dag}b_{2\mathbf{k}\sigma}+
b_{2\mathbf{k}\sigma}^{\dag}a_{1\mathbf{k}\sigma}] ,\label{eq:03}
\end{eqnarray}
to obtain it's energy bands which are
\begin{eqnarray}
E_{\lambda
}^{\nu}=\nu(\sqrt{|\phi(\mathbf{k})|^{2}+(\frac{\gamma}{2})^{2}}+(-1)^{\lambda}\frac{\gamma}{2}),
\label{eq:04}
\end{eqnarray}
where $\lambda=1,2$ are the energy bands number and $\nu=+(-)$
indicates the conduction(valance) energy bands respectively.
$\phi(\mathbf{k})=\sum_{i=1}^{^{3}}e^{i\mathbf{k}.\vec{\delta}_{i}}$
and $\vec{\delta}_{1}=a(\sqrt{3}\hat{x}/2+\hat{y}/2)$,
$\vec{\delta}_{2}=a(-\sqrt{3}\hat{x}/2+\hat{y}/2)$ and
$\vec{\delta}_{3}=-a\hat{y}$ which are the nearest neighbor
vectors. One can expand $|\phi(\mathbf{k})|$ around Dirac points
($\mathbf{K}$ or $\mathbf{K}^{'}$) for
$|\mathbf{q}|\ll|\mathbf{K}|$ (where
$\mathbf{k}=\mathbf{q}+\mathbf{K}$) to obtain the low energy bands
of BLG \cite{Castro Neto1}. In this limit
$|\phi(\mathbf{k})|=v_{F}q$ where $v_{F}=3ta/2$ $(\approx 10^{6}$
m/s) is the Fermi velocity of the Dirac electrons.

The Hamiltonian of the impurity is
\begin{eqnarray}
H_{imp}=\varepsilon_{0}\sum_{\sigma}f_{\sigma}^{\dag}f_{\sigma}+
Un_{\uparrow}n_{\downarrow},\label{eq:05}
\end{eqnarray}
where $\varepsilon_{0}$ is the energy of the impurity state when
it is occupied by one electron and $U$ is the Coulomb energy for
double occupancy of the impurity state.
$n_{\sigma}=f_{\sigma}^{\dag}f_{\sigma}$ is the occupation number
and $f_{\sigma}^{\dag}(f_{\sigma})$ is the creation (annihilation)
operator of an electron with spin $\sigma$ at the impurity state.
Following Anderson\cite{Anderson} we use the mean field
approximation to decouple the two-body interaction term as
$\sum_{\sigma}\langle n_{-\sigma}\rangle
f_{\sigma}^{\dag}f_{\sigma}-\langle n_{\uparrow}\rangle \langle
n_{\downarrow}\rangle$. Hence we can rewrite the Hamiltonian of
the impurity as $
\sum_{\sigma}\varepsilon_{\sigma}f_{\sigma}^{\dag}f_{\sigma},\label{eq:11}
$ where $\varepsilon_{\sigma}=\varepsilon_{0}-U\langle
n_{-\sigma}\rangle$ is the renormalized energy of the impurity
state.

The localized state of the impurity can hybridize with $\pi$ band
of BLG at the adsorb location via following Hamiltonian
\begin{eqnarray}
H_{V}=V\sum_{\sigma}(f_{\sigma}^{\dag}a_{\sigma}(0)+a_{\sigma}^{\dag}(0)f_{\sigma}).\label{eq:06}
\end{eqnarray}
Here the impurity is adsorbed on the $A$ sublattice of site 0 and $V$ is the
hybridization strength. Depending on the strength of the
hybridization and the on-site coulomb energy, two cases are
possible. For the case $n_{\uparrow}\neq n_{\downarrow}$ the local moment forms
but for the case $n_{\uparrow}=n_{\downarrow}$ no local local moment.
The occupation number of a spin impurity state
at zero temperature is given by
\begin{eqnarray}
n_{\sigma}=\int_{-\infty}^{\mu}d\omega \rho_{\sigma}^{imp}(\omega)
,\label{eq:07}
\end{eqnarray}
where
$\rho_{\sigma}^{imp}(\omega)=-\frac{1}{\pi}ImG^{imp,R}_{\sigma}(\omega)$.
We use the equation of motion technique to write the retarded
Green's function of the impurity, $G^{imp,R}_{\sigma}(\omega)$, as
\begin{eqnarray}
G^{imp,R}_{\sigma}(\omega)=[\omega-\varepsilon_{\sigma}-\Sigma^{R}_{imp}(\omega)+i0^{+}]^{-1}
,\label{eq:08}
\end{eqnarray}
in which the retarded self-energy of the impurity is
\begin{eqnarray}
\Sigma^{R}_{imp}(\omega)=V^{2}G^{0R}_{\alpha\alpha,\sigma}(\omega)=\frac{V^{2}}
{\sqrt{N}}\sum_{\mathbf{q}}G^{0R}_{\alpha\alpha,\sigma}(\mathbf{q},\omega)
,\label{eq:09}
\end{eqnarray}
where $\alpha=A$ ($\alpha=B$) when the impurity is
adsorbed on $A$
($B$) sublattice of BLG and
$G^{0R}_{\alpha\alpha,\sigma}(\mathbf{q},\omega)$ is the
clean Green's function of BLG at $\alpha$ sublattice.
After integration over momentum we found the Green's functions as
\begin{eqnarray}
G^{0R}_{AA,\sigma}(\omega)=-\frac{1}{2D^{2}}[\omega\ln|\frac{(\omega-\gamma/2)^{2}-D^{2}}{\omega(\omega-\gamma)}|+
\omega\ln|\frac{(\omega+\gamma/2)^{2}-D^{2}}{\omega(\omega+\gamma)}|]\nonumber
\\-i\frac{\pi}{2D^{2}}[|\omega|\theta(\gamma-|\omega|)+2|\omega|\theta(|\omega|-\gamma)]\theta(|\omega|-D)
,\label{eq:10}
\end{eqnarray}
and
\begin{eqnarray}
G^{0R}_{BB,\sigma}(\omega)=-\frac{1}{2D^{2}}[(\omega-\gamma)\ln|\frac{(\omega-\gamma/2)^{2}-D^{2}}
{\omega(\omega-\gamma)}|+
(\omega+\gamma)\ln|\frac{(\omega+\gamma/2)^{2}-D^{2}}{\omega(\omega+\gamma)}|]\nonumber
\\-i\frac{\pi}{2D^{2}}[(|\omega|+\gamma)\theta(\gamma-|\omega|)+2|\omega|\theta(|\omega|-\gamma)]
\theta(|\omega|-D),\label{eq:11}
\end{eqnarray}
where $D$ is the high-energy cutoff of BLG bandwidth and
$\theta(x)$ is the step function. Eqs. (\ref{eq:07})-(\ref{eq:11})
construct a closed set of equations which should be solved
selfconsistently to obtain the occupation number. Our numerical
results are presented in the next section.

\section{Numerical Results}
\label{sec:4}

The local magnetic moment exists when the occupation number of two
spins at the impurity level are different, namely whenever
$n_{\uparrow}\neq n_{\downarrow}$. The occupation number for a
giving spin, $n_{\sigma}$, can be calculated self-consistently
from Eqs. (\ref{eq:07})-(\ref{eq:11}). We use following
dimensionless parameters, $x=\Delta D/U$ and
$y=(\mu-\varepsilon_{0})/U$ with $\Delta=\pi V^{2}/D^{2}$, in the
reminder of this paper.

Fig. \ref{fig:02} presents our results for the curves of the
boundary separating the magnetic and non-magnetic phase of the
impurity state. These curves are corresponding to different values
of the interlayer tunnelling energies, $\gamma=0$, $\gamma=0.175$
eV and $\gamma=0.4$ eV, when the impurity is adsorbed on $A$
sublattice (left panel) or on $B$ sublattice (right panel). The
other parameters are $\varepsilon_{0}=0.2$ eV, $V=1.0$ eV and
$D\approx7$ eV. We considered all possible locations for the
impurity adsorption on BLG sublattices. Our results for BLG case
when the impurity is adsorbed on $B$ sublattice and for SLG limit
($\gamma=0$) are in agreement with previous
works\cite{Uchoa,Ding}. Also we see that by increasing the
interlayer tunnelling energy (transforming from two separated SLG
to BLG case) for the impurity adsorbed on $A$ sublattice the size
of the magnetic region increases while when the impurity is
adsorbed on $B$ sublattice it decreases. Note that the local
density of states (LDOS) at $A$ and $B$ sublattices of
BLG\cite{ZWang} are
\begin{eqnarray}
N_{A}(\omega)=\frac{1}{2D^{2}}[|\omega|\theta(\gamma-|\omega|)+2|\omega|\theta(|\omega|-\gamma)]
\theta(|\omega|-D),\label{eq:12}
\end{eqnarray}
and
\begin{eqnarray}
N_{B}(\omega)=\frac{1}{2D^{2}}[(|\omega|+\gamma)\theta(\gamma-|\omega|)+2|\omega|\theta(|\omega|-\gamma)]
\theta(|\omega|-D),\label{eq:13}
\end{eqnarray}
respectively. We see that by increasing the interlayer tunnelling
energy, the local density of the low energy states at $A$
sublattice decreases, so the hybridization of the impurity level
with carbon atom located at $A$ sublattice becomes weaker with
respect to that in SLG. This leads to easier formation of the
localized magnetic moment on the impurity adsorbed on $A$
sublattice. So the size of the magnetic region increases. But for
the impurity adsorbed on $B$ sublattice, increasing the interlayer
tunnelling energy increases the local density of the low energy
states. This leads to enhancement of the hybridization with
impurity state. So the magnetic moment formation region decreases.

Furthermore when the impurity is adsorbed on $A$ sublattice, the
magnetic boundary crosses the line $y=0$. This is due to the large
broadening of the impurity level, as the local moment can form
even when $\varepsilon_{0}$ is above the Fermi energy. But when
the impurity is adsorbed on $B$ sublattice this feature disappears approximately.
 This could be explained by the small broadening of
impurity level which is due to the nonzero amount of the LDOS at
around the impurity level.

Fig. \ref{fig:03} shows effects of
$\varepsilon_{0}$ and $V$ variation on the size of the magnetic region. We
see that as $\varepsilon_{0} \rightarrow 0$ the local
moment formation becomes more possible so the size of the magnetic
region grows (for impurities adsorbed on both A and B
sublattices). This can be explained by this fact that the LDOS
around the impurity level is suppressed which allows for easy
formation of the local moment. Furthermore when the impurity is
adsorbed on A sublattice the size of the magnetic region increases
more, because at low energy states, $\omega \rightarrow 0$, the LDOS on A sublattice decreases
faster than that on B sublattice. Also as we expect, by increasing
the hybridization strength the size of the magnetic region
decreases.

Similar to SLG, in the BLG the magnetization of the impurity can
be controlled by applying an electric field via a back gate which
can change the chemical potential. This allows for the
potentiality of BLG for spintronics. This can be clarified by
considering the chemical potential dependence of the occupation
number and susceptibility of the magnetic impurity. The magnetic
susceptibility of the impurity is defined as
$\chi=\mu_{B}\sum_{\sigma}\sigma(dn_{\sigma}/dB)_{B=0}$ in the
zero magnetic field limit. This can be rewritten as

\begin{eqnarray}
\chi=-\mu_{B}^{2}\sum_{\sigma}\frac{d\langle
n_{\sigma}\rangle}{d\varepsilon_{\sigma}}\frac{1-U\frac{d\langle
n_{-\sigma}\rangle}{d\varepsilon_{-\sigma}}}{1-U^{2}\frac{d\langle
n_{-\sigma}\rangle}{d\varepsilon_{-\sigma}}\frac{d\langle
n_{\sigma}\rangle}{d\varepsilon_{\sigma}}},\label{eq:12}
\end{eqnarray}
where
$\varepsilon_{\sigma}=\varepsilon_{0}-\sigma\mu_{B}B+Un_{-\sigma}$
is the energy of the impurity spin state in the presence of the
magnetic field which tends to zero. The impurity magnetizes when a
bubble shape exists in the occupation number curve. At the edges
of this magnetic bubble the corresponding susceptibility has two
peaks which show the strength of the magnetic transition.

In Fig. \ref{fig:04} we plotted $n_{\sigma}(\mu)$ and $\chi(\mu)$
of a magnetic impurity adsorbed on A sublattice (left panels) and
B sublattice (right panels) for different values of the on-site
coulomb energies, $U=96$ meV (dotted-dashed curves), $U=48$ meV
(dashed curves) and $U=24$ meV (solid curves). The other
parameters are considered as $\varepsilon_{0}=0.2$ eV, $V=1$ eV
and $\gamma=0.4$ eV. We see that one can control amount of the
local moment of the magnetic impurity adsorbed on both A and B
sublattices via varying the chemical potential. Furthermore Figs.
\ref{fig:04} shows that controlling the local magnetic moment on A
sublattice is possible for wide range of the on-site coulomb
energies.

It has been reported in the recent considerations that the
magnetic coupling between magnetic moment of adatoms adsorbed on
same (different) sublattices is ferromagnetic (antiferromagnetic)
in both single layer graphene (SLG) \cite{Saremi} and bilayer
graphene (BLG) \cite{Jiang,Parhizgar}. Furthermore we showed that
the local moment strengths of magnetic adatoms on different
sublattices of BLG are not equal. Hence in BLG even when magnetic
adatoms are distributed randomly on different sublattices we have
net local moment. While in SLG, because the local moment strengths
of magnetic adatoms adsorbed on different siblattices are
equal\cite{Saremi}, for random distribution of magnetic adatoms
there is not net local moment. These features allow for using BLG
as spin switcher in the spintronics devices with higher potential
than SLG.

This figure also sustains our previous results about more
possibility of the impurity magnetization on A sublattice in
contrast to B sublattice. For the impurity adsorbed on A
sublattice with mentioned parameters, at a small on-site coulomb
energy about $U=96$ meV a strong magnetic moment of $\sim0.8
\mu_{B}$ forms in the entire magnetic region approximately. By
decreasing the on-site coulomb energy, size of the magnetic region
decreases and the magnetic transition phase becomes very sharp
(Fig. \ref{fig:04}(a) and \ref{fig:04}(b)), but even at $U=24$ meV
a local magnetic moment of $\sim0.3 \mu_{B}$ forms. The impurity
magnetization for a such small on-site coulomb energy neither in
normal metal \cite{Mahan} nor in SLG \cite{Uchoa} has not been
reported. While when the impurity is adsorbed on B sublattice the
impurity magnetization is possible for on-site coulomb energies
which are nearly thrice larger. Also we see that by decreasing $U$
the size of the magnetic bubble rapidly diminishes.

\section{Summary and conclusions}
\label{sec:5}

In summary, we considered the local magnetic moment formation of a magnetic
impurity adsorbed on a BLG. We found different features for the
magnetization of the impurity adsorbed on A and B sublattices.

We showed that the magnetic impurity adsorbed on A sublattice can
magnetize even at very small on-site coulomb energies. This is not
reported neither in normal metal \cite{Mahan} nor in SLG
\cite{Uchoa}. This is due to the very low LDOS at A sublattice
that decreases the effect of the hybridization with the impurity
level and allow for easy formation of the local moment. Also we
found that the local moment forms even when the energy of bare
impurity level is above the Fermi energy. This can be explained by
large broadening of the impurity level.

But when the impurity is adsorbed on B sublattice, due to the
large local density at low energy states the hybridization with
the impurity level is enhanced. This limits the impurity
magnetization on B sublattice in comparison with that on A
sublattice of BLG and also with that on all sublattices of SLG.
Also due to the small broadening of the impurity level, the
impurity magnetizes approximately only when the bare impurity
level is below the Fermi energy.

Finally we showed that by varying the chemical potential via an
external electric field one can control the local moment of the
magnetic impurity adsorbed on both A and B sublattices. This
feature besides inhomogeneity of the adatom's magnetization in BLG
allow for using BLG as spin switcher in spintronics devices.

%

%
%
%
\newpage
\begin{figure}
\begin{center}
\includegraphics[width=12cm,angle=270]{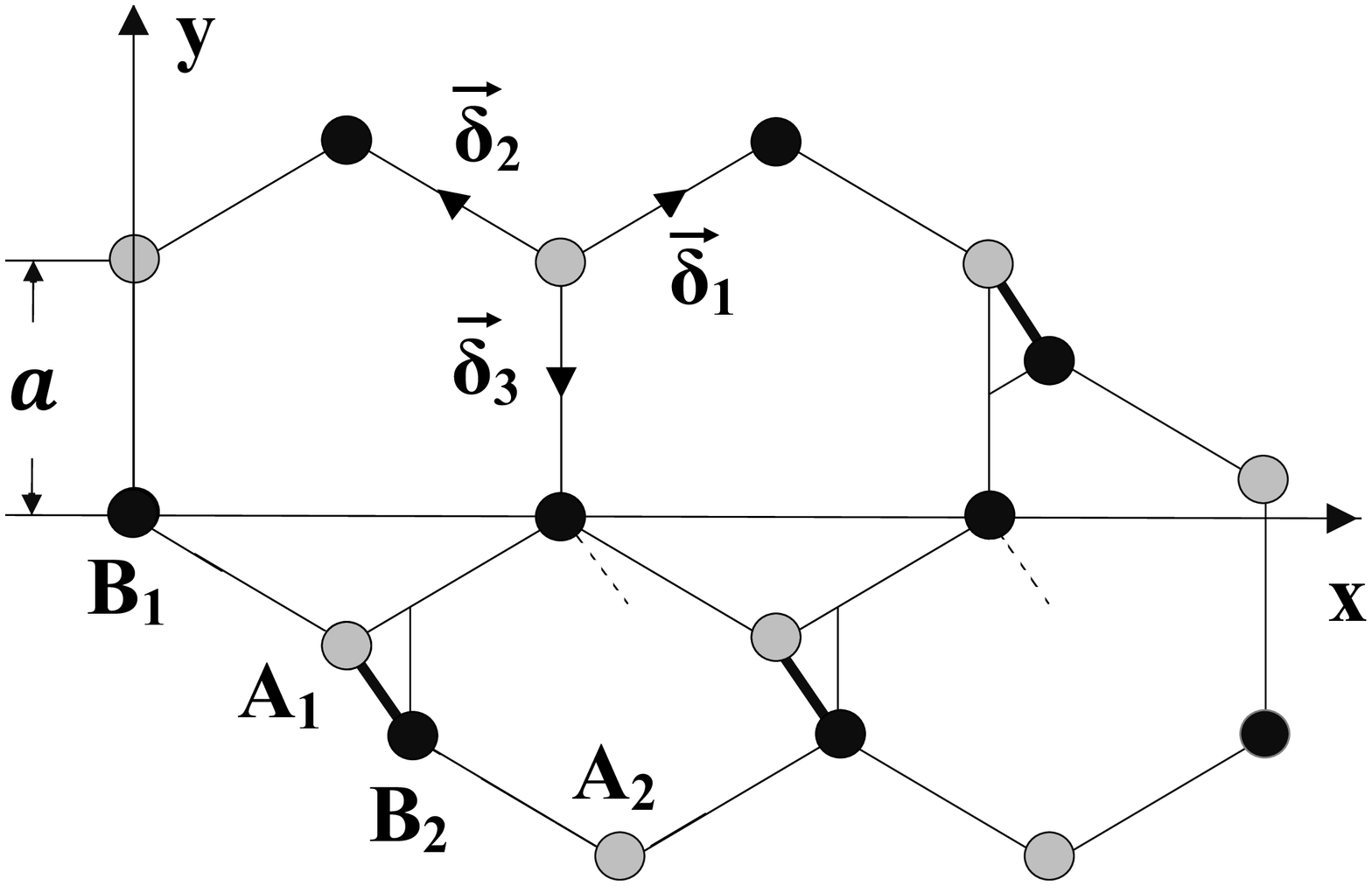}
\caption{Lattice structure of BLG.
$\vec{\delta}_{1}=a(\sqrt{3}\hat{x}/2+\hat{y}/2)$,
$\vec{\delta}_{2}=a(-\sqrt{3}\hat{x}/2+\hat{y}/2)$ and
$\vec{\delta}_{3}=-a\hat{y}$ are three vectors that are drown from
connects $A_{1}$ sublattice to it's nearest
neighbors.}\label{fig:01}
\end{center}
\end{figure}
\begin{figure}
\begin{center}
\includegraphics[width=15cm,angle=0]{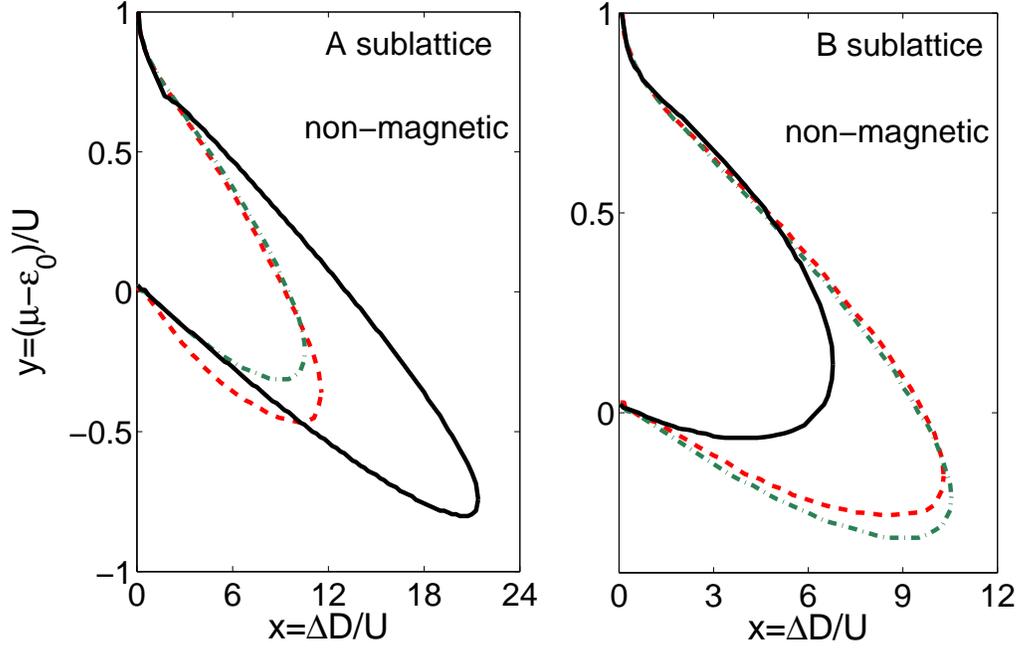}
\caption{The boundary between magnetic and non-magnetic states of
an impurity adsorbed on A sublattice (Left panel) and B sublattice
(Right panel) in a bilayer graphene lattice for different
interlayer hopping, $\gamma_{1}=0.4$(Solid curves),
$\gamma_{1}=0.185 eV $(dashed curves) and $\gamma_{1}=0.0
eV$(doted-dashed curves). The other parameters are
$\varepsilon_{0}=0.2 eV$, $V=1 eV$. }\label{fig:02}
\end{center}
\end{figure}
\begin{figure}
\begin{center}
\includegraphics[width=15cm,angle=0]{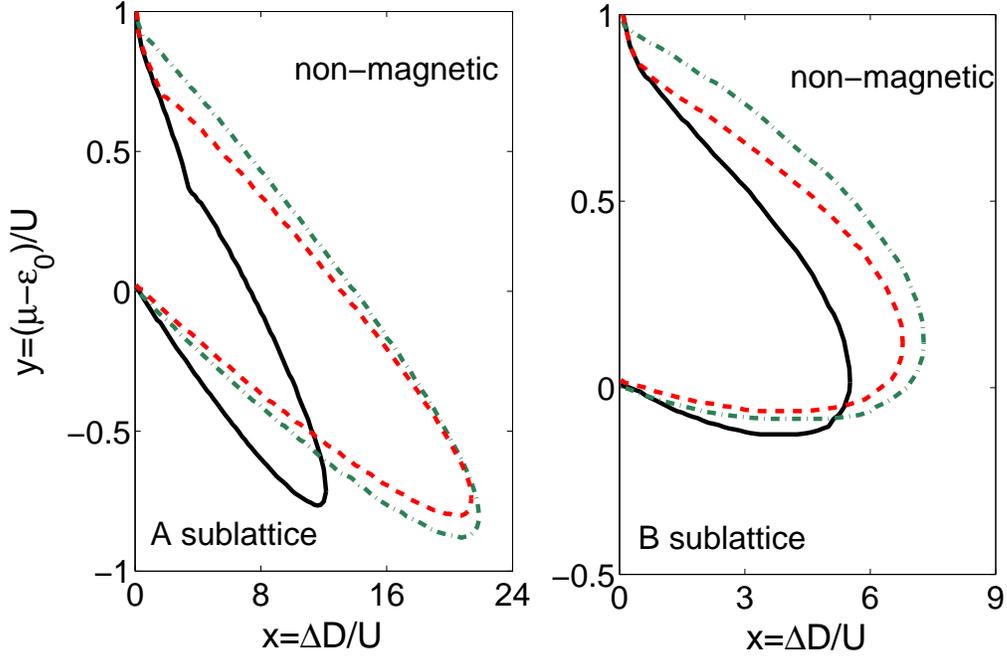}
\caption{The boundary between magnetic and non-magnetic states of
an impurity adsorbed on A sublattice (Left panel) and B sublattice
(Right panel) in a bilayer graphene lattice with $\gamma=0.4eV$.
Solid curves: $\varepsilon_{0}=0.35eV$, $V=1eV$; dashed curves:
$\varepsilon_{0}=0.2eV$, $V=1eV$; doted-dashed:
$\varepsilon_{0}=0.2eV$, $V=0.5eV$.}\label{fig:03}
\end{center}
\end{figure}
\begin{figure}
\begin{center}
\includegraphics[width=15cm,angle=0]{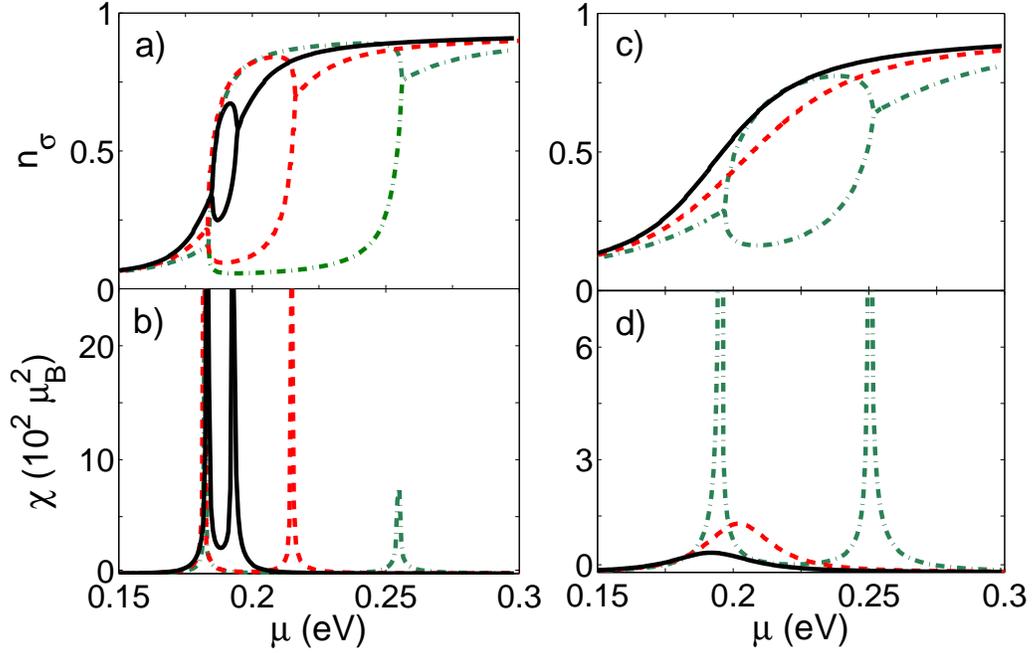}
\caption{$n_{\sigma}(\mu)$ and $\chi(\mu)$ curves of an magnetic
impurity with $\varepsilon_{0}=0.2eV$ and $V=1eV$ adsorbed on A
sublattice (left panels) and B sublattice (right panels). Dotted-dashed: $U=0.096eV$; dashed curves: $U=0.048eV$; solid
curves: $U=0.024eV$.}\label{fig:04}
\end{center}
\end{figure}
\end{document}